\documentstyle[12pt]{article}
\begin{document}
\parskip=0pt
\parindent=123mm

{\hskip -10mm} TECHNION-PH-97-4\par

\bigskip
\bigskip
\begin{center}
\parskip=10pt
{\Large\bf Two-photon exclusive decays}\\[0.3cm]
{\Large\bf $B_s \to \eta (\eta') \gamma\gamma$ 
and $B \to K \gamma\gamma$}\\[4cm]
{\large Paul Singer and Da-Xin Zhang}\\[0.5cm]
Department of Physics,
Technion -- Israel Institute of Technology,
Haifa 32000, Israel\\[3cm]
{\bf ABSTRACT}\\
\end{center}
\parindent=23pt

\noindent
The exclusive decay modes 
$B \to K \gamma\gamma$ and
$B_s \to \eta (\eta') \gamma\gamma$ 
are shown to have significant branching ratios of
approximately $0.5\times 10^{-7}$.
This first calculation of these modes employs a model
based on a cascade transition
$B\to V\gamma\to P\gamma\gamma$
for estimating the long-distance contribution
and the process $b\to s\gamma\gamma$ for the
short distance one.

\vskip 2cm
\noindent {\it PACS number(s):}
12.40.Vv,  12.20.He, 12.15.Mm

\newpage

The investigation of flavor-changing weak radiative
transitions of $b$-quark has emerged during
the last decade as a most fruitful field of research,
both experimentally and theoretically.
There are already two measurements by CLEO of 
transitions in which the electromagnetic penguin
$b\to s\gamma$ plays a dominant role,
of the inclusive decay transition
$BR(B\to X_s\gamma)=(2.32\pm 0.57\pm 0.35)\times 10^{-4}$\cite{1}
and of an exclusive transition
$BR(B\to K^*\gamma)=(4.5\pm 1.5\pm 0.9)\times 10^{-5}$\cite{2}.
These measurements confirm within the existing accuracy 
the Standard Model (SM) prediction,
including QCD corrections\cite{3},
for these modes.
In particular,
with the inclusion of next to leading order QCD  corrections\cite{4}
one arrives at the SM theoretical prediction 
$BR(B\to X_s\gamma)^{th}=(3.28\pm 0.33)\times 10^{-4}$.
Consequently,
the study of these decays is also a good
testing ground for new physics\cite{5}.
Concerning the long-distance contributions,
we remark that various theoretical treatments\cite{6}
agree that such  contributions are small compared to the
short-distance electromagnetic penguin $b\to s\gamma$,
most probably amounting to less than $(5-8)\%$ of the rate.

The two-photon decays of $B$-mesons,
related to the quark transition $b\to s\gamma\gamma$,
are also of considerable interest\cite{7,8,9}.
Considering only the contributions of the quark transition 
 $b\to s\gamma\gamma$,
without QCD-corrections,
these authors found a fairly sizable rate within SM for the
two-photon decay of $B_s$, 
$BR(B_s\to \gamma\gamma)\simeq (1.5-3)\times 10^{-7}$,
which should be measurable at future $B$-machines.

In a recent paper\cite{10},
a detailed study of the inclusive rare process $B\to X_s\gamma\gamma$
has been undertaken in terms\cite{11} of the quark level transition
 $b\to s\gamma\gamma$,
both in SM and in two versions of the Two Higgs Doublet model.
They find in SM a branching ratio
of $\sim 1\times 10^{-7}$ for decay into hard photons,
of energies above $100$MeV each.

In the present paper we propose a model for calculating
the exclusive decays of $B$-mesons into one
pseudoscalar meson ($P$) and two photons,
concentrating on the modes 
$B_s \to \eta \gamma\gamma$,
 $B_s \to \eta' \gamma\gamma$,
and $B^{+,0} \to K^{+,0} \gamma\gamma$.
The calculation includes both
short-distance and long-distance contributions.
Our results reveal two interesting features of these modes:
firstly, they appear to contribute a large proportion of the
inclusive rate,
as estimated in Ref. \cite{10},
and secondly,
the short- and long-distance contributions are of comparable magnitude.
Decays of type $B(B_s)\to V \gamma\gamma$,
involve different features and will be discussed 
in a separate publication\cite{12}.

We start with the consideration of the  $b\to s\gamma\gamma$ transition,
which is at the core of our treatment.
The basic amplitude at the quark level
is given \cite{7,8,9} by the expression:
\begin{eqnarray}
A(b\rightarrow s\gamma\gamma)=-i\frac{\alpha_e G_F}{\sqrt{2}\pi}
\epsilon_{\mu}(k_1)\epsilon_{\nu}(k_2)\bar u(p_s)T^{\mu\nu}u(p_b),
\label{bsrr1}
\end{eqnarray}
where
\begin{eqnarray}
T^{\mu\nu}=\sum_{i=u,c,t}\lambda_i
T_i^{\mu\nu}=\lambda_u(T^{\mu\nu}_u-T^{\mu\nu}_c)
+\lambda_t(T^{\mu\nu}_t-T^{\mu\nu}_c),
\label{tmunu}
\end{eqnarray}
and $\lambda_i=V_{ib}V^*_{is}$.
These tensors $T_i^{\mu\nu}$
are divided into the one-particle-irreducible (1PI) 
and the one-particle-reducible (1PR) parts.
The 1PI part is induced by diagrams with both photons
coming from the quark-W boson loop,
while the 1PR diagrams have one photon
emitted from the external $b$ or $s$ quark line.

For concreteness,
we present the details of our approach by treating the 
$B_s\to \eta\gamma\gamma$ decay.
Afterwards,
we shall refer to three additional modes,
$B_s\to \eta'\gamma\gamma$ and $B^{+,0} \to K^{+,0} \gamma\gamma$.
The 1PI contribution
 constituting the short-distance part of the exclusive decay,
can be calculated directly.
The long-distance part of it may be represented by
$B_s\to\eta_c\eta\to\gamma\gamma\eta$
and we estimate this process to be less important.
On the other hand,
the 1PR part needs to be performed at the hadronic level and will
provide the long-distance part of the exclusive transition.
For this part,
we construct a model which uses vector-meson-dominance (VMD),
so that the final  $\eta\gamma\gamma$ state is realized via the cascade decay
$B_s\to\phi\gamma, \phi\to\eta\gamma$.
Both these intermediate transitions have sizable strength,
which is essentially under control
($\phi\to\eta\gamma$ is measured and $B_s\to\phi\gamma$
is related to the observed $B\to K^*\gamma$ decay).
Accordingly,
we consider our model to be a  reliable tool for
the estimation of $B(B_s)\to P\gamma\gamma$ decays.
We also made rough estimates for other possible
cascade mechanisms,
like  $B_s\to B_s^*\gamma\to \eta\gamma\gamma$,
and we find this contribution to be significantly smaller.

The 1PI part of the transition $b\rightarrow s\gamma\gamma$
with on-shell photons with momenta $k_1, k_2$
 is given by\cite{9,10}
\begin{eqnarray}
T^{\mu\nu}_{i,\rm{1PI}}%&=&\frac{8}{9}\delta_3(z_i)I^{\mu\nu},\\
%&=&
=\frac{8}{9}\delta_3(z_i)
\left[
i\epsilon^{\mu\nu\xi\alpha}\gamma_{\alpha}\gamma_L (k_1-k_2)_{\xi}+
i\frac{k_{1\xi}k_{2\eta}}{k_1\cdot k_2}
(\epsilon^{\mu\xi\eta\alpha}k_1^{\nu}-
\epsilon^{\nu\xi\eta\alpha}k_2^{\mu})\gamma_{\alpha}\gamma_L\right],
\end{eqnarray}
 where $\gamma_L=(1-\gamma_5)/2$, $z_i=2k_1\cdot k_2/m_i^2$, 
and\cite{13}
\begin{eqnarray}
\delta_3(z_i)=1+\frac{2}{z_i}\int_0^1\frac{du}{u}
\log\left[1-z_iu(1-u)\right].
\label{delta3}
\end{eqnarray}
This function $\delta_3(z_i)$ has been explored in details
in \cite{9}.

To calculate the decay amplitude of the 1PI part,
we need the hadronic matrix element
\begin{eqnarray}
\langle\eta|\bar s \gamma_{\alpha}\gamma_L b|\bar B_s\rangle
=\frac{1}{2}f_+(q^2)(P_{B_s}+p_\eta)_{\alpha}+
\frac{1}{2}f_-(q^2)(P_{B_s}-p_\eta)_{\alpha},
\end{eqnarray}
$q^2=(P_{B_s}-p_\eta)^2$.
For  the formfactors $f_\pm$ at zero recoil we use the
 chiral perturbation theory\cite{14} which gives
\begin{eqnarray}
f_+((m_{B_s}-m_\eta)^2)= -f_-((m_{B_s}-m_\eta)^2)
=-\frac{f_{B_s}}{f_\eta}\frac{g_{B_s^*B_s\eta}m_{B_s}}{\Delta+m_\eta}
\sqrt{\frac{2}{3}},
\label{formfactor}
\end{eqnarray}
where $\Delta=m_{B_s^*}-m_{B_s}$.
Away from the zero recoil point a monopole
behavior will be used:
\begin{eqnarray}
f_\pm(q^2)=f_\pm((m_{B_s}-m_\eta)^2)
\frac{1-(m_{B_s}-m_\eta)^2/m_{B_s^*}^2}{1-q^2/m_{B_s^*}^2}.
\end{eqnarray}
The result for the short-distance part of $B_s\to\eta\gamma\gamma$
is thus given by
\begin{eqnarray}
A^{1PI}(B_s\to\eta\gamma\gamma)&=&
\frac{\alpha_eG_F}{\sqrt{2}\pi}\frac{8}{9}
[\lambda_u(\delta_3(z_u)-\delta_3(z_c))
+\lambda_t(\delta_3(z_t)-\delta_3(z_c)) ]
f_+(q^2)\varepsilon_{\mu\nu\alpha\beta}\\
\nonumber
&&\cdot [\epsilon_1^\mu\epsilon_2^\nu(k_1-k_2)^\alpha p_\eta^\beta
+
(\epsilon_2\cdot k_1 \epsilon_1^\mu {k_1}^\nu {k_2}^\alpha p_\eta^\beta
 -\epsilon_1\cdot k_2 \epsilon_2^\mu {k_1}^\nu {k_2}^\alpha p_\eta^\beta)
/(k_1\cdot k_2)].
\label{a1pi}
\end{eqnarray}

We turn now to the reducible part of the amplitude,
which we assume to proceed via $B_s\to\phi\gamma\to\eta\gamma\gamma$.
The amplitude for $B_s\to \phi\gamma$ is\cite{15}
\begin{eqnarray}
A(B_s\to \phi\gamma(k_1, \epsilon_1))&=&
\displaystyle\frac{G_Fem_b}{2\sqrt{2}\pi^2}
V_{tb}V^*_{ts}C_7^{eff}
[T_1(0)i\varepsilon^{\alpha\beta\mu\nu}
\epsilon^\phi_\alpha\epsilon_{1\beta} p_{B_s\mu}k_{1\nu}\\
\nonumber
&&
+T_2(0)((\epsilon^\phi\cdot\epsilon_1)(p_{B_s}\cdot k_1)
-(\epsilon^\phi\cdot k_1)(p_{B_s}\cdot\epsilon_1))],
\label{e10}
\end{eqnarray}
where $C_7^{eff}=0.65$, and $T_1(0)=T_2(0)=0.115$.
This decay is driven by the $b\to s\gamma$ transition,
defined by
\begin{eqnarray}
A(b\to s\gamma)&=&
\frac{2G_F}{\sqrt{2}}\lambda_tC_7^{eff}(m_b)O_7,
~~~ {\rm with}~~ O_7=\frac{em_b}{16\pi^2}F_{\mu\nu}
\bar s\sigma^{\mu\nu}\frac{1+\gamma_5}{2} b.
\label{e11}
\end{eqnarray}
We note that from Eqs. (9) and (10) one predicts\cite{15}
a branching ratio for $B_s\to\phi\gamma$  comparable to
that of $B\to K^*\gamma$,
which is indeed expected.

The amplitude for $\phi\to\eta\gamma$ is
\begin{eqnarray}
A(\phi\to\eta\gamma(k_2, \epsilon_2))=c_{\phi\eta\gamma}
\varepsilon^{\alpha\beta\mu\nu}
\epsilon^\phi_\alpha\epsilon_{2\beta} p_{\eta\mu} k_{2\nu},
\end{eqnarray}
where $|c_{\phi\eta\gamma}|=0.21{\rm GeV}^{-1}$  is determined from the
partial decay width.
Thus the 1PR part of the amplitude turns out to be in the VMD model
\begin{eqnarray}
A^{1PR}(B_s\to\eta\gamma\gamma)&=&c_{\phi\eta\gamma}
\frac{G_Fem_b}{2\sqrt{2}\pi^2}\lambda_t C_7^{eff}
[T_1(0)i\varepsilon_{\alpha\beta\mu\nu}
\epsilon_1^\beta P_{B_s}^\mu k_1^\nu
+T_2(0)(\epsilon_{1\alpha} P_{B_s}\cdot k_1
-k_{1\alpha} P_{B_s}\cdot \epsilon_1)]\nonumber\\
\nonumber
&&\cdot \varepsilon^{\alpha\gamma\delta\rho}
{\epsilon_2}_\gamma {p_\eta}_\delta {k_2}_\rho
\frac{1}{(p_\eta+{k_2})^2-m_\phi^2+im_\phi\Gamma_\phi}\\
&+&(k_1\leftrightarrow k_2, \epsilon_1\leftrightarrow \epsilon_2).
\label{a1pr}
\end{eqnarray}
This treatment of the 1PR amplitude is
consistent with the decomposition theorem\cite{7,16}.

We turn now to extend our calculation to include also
the $B_s\to\eta'\gamma\gamma$ mode.
For this purpose,
if we use the nonet symmetry between the octet and the singlet
 light pseudoscalars,
we need to consider the $\eta - \eta'$ mixing,
which is defined by
\begin{eqnarray}
\eta &=& \eta_8 {\rm cos}\theta_p - \eta_0 {\rm sin}\theta_p\\
\nonumber
\eta' &=& \eta_8 {\rm sin}\theta_p + \eta_0 {\rm cos}\theta_p,
\end{eqnarray}
and the factor $\sqrt{2/3}$ in Eq. (6)
should be repaced by 
$\sqrt{2/3} {\rm cos}\theta_p+\sqrt{1/3} {\rm sin}\theta_p$
for the $B_s\to \eta$ transition,
and by 
$\sqrt{2/3} {\rm sin}\theta_p-\sqrt{1/3} {\rm cos}\theta_p$
for the $B_s\to \eta'$ transition.
Numerically $\theta_p \sim -20^\circ$ will be used.
Now, the physical masses of $\eta$ or $\eta'$ will be used
in Eq. (6).
In this way,
Eq. (8) is extended to cover the decays to $\eta$ and $\eta'$,
with the appropriate replacements in (6) and (7).

For the long-distance part of $B_s\to\eta'\gamma\gamma$,
we use again the nonet symmetry, 
under which the coupling $c_{\phi  \eta'  \gamma}$ for 
$\phi \to \eta'  \gamma$ is related
to $c_{\phi\eta\gamma}$ by\cite{17}
\begin{eqnarray}
|c_{\phi  \eta'  \gamma}/c_{\phi\eta\gamma}|
=({\rm cos}\theta_p - \sqrt{2} {\rm sin}\theta_p)
/({\rm sin}\theta_p + \sqrt{2} {\rm cos}\theta_p),
\end{eqnarray}
which gives $|c_{\phi \eta'  \gamma}|=0.30{\rm GeV}^{-1}$
if $\theta_p=-20^\circ $ is used.
A completely similar calculation is carried out for $B \to K \gamma\gamma$,
in which case the long-distance part is due to
 $(B^{+,0} \to K^{*+,0} \gamma)+(K^{*+,0}\to K^{+,0}\gamma)$.
Here,
the required  branching ratios are known for both transitions\cite{18}.

In presenting our results, we recognize that unfortunately
we cannot fix the relative phase between
 the 1PR and the 1PI amplitudes.
This makes the phase of the interference 
effects in the decay rates undetermined.
The reason can be seen from the 
 amplitudes in  (9) and in (12).
There are too many   Lorentz  structures involved, 
which make the analogue Argand plot analysis\cite{19} invalid.
Thus we will present the results 
for the 1PI, the 1PR and the interference contributions separately .
The total decay rates are then 
$\Gamma_{1PI}+\Gamma_{1PR}\pm \Gamma_{inter}$.
As a result,
we also refrain from presenting differential decay distributions
in various variables.

We follow now the authors of Ref.  \cite{10} by imposing
several cut conditions on the decays considered,
which were enforced on 
the quark level transition $b\to s\gamma\gamma$ in Ref. \cite{10}.
First, 
hard photons with energy larger than 100MeV are chosen.
Second,
the invariant mass squared for any two final particles is
demanded to be larger than $cm_B^2$ or $cm_{B_s}^2$,
 with $c=$0.01 or 0.02.
Third,
all the angles between two of the  final particles are taken to be
 larger than $20^\circ$.
In addition,
we wish to exclude from the decay the region
of $\eta\gamma, \eta'\gamma$ and $K\gamma$ 
which is close to the resonance peak.
Thus,
we demand that the invariant mass  of any $\eta\gamma, \eta'\gamma$ pair
deviates from the $\phi$ resonance by more than $\pm 50$(or $100$)MeV,
while for $K\gamma$ the requirement extends to $300$(or $500$)MeV
(the  width of $\phi$ is $4.43$MeV and of $K^*$ is $50$MeV).
We  also used a modified form of the new requirement 
by demanding that the invariant mass  of any $\eta(\eta')\gamma$ pair
 be larger than $1.1$GeV or $1.2$GeV,
and of $K\gamma$ be larger than $1.2$GeV or $1.4$GeV.
Using the same mass parameters and CKM matrix elements as in \cite{10},
and $f_\eta=0.13$GeV, $f_{\eta'}=0.11$GeV, $f_K=0.16$GeV\cite{20},
$g_{B_s^*B_s\eta}=0.5$,
the results are given in Table 1.
In the calculation we approximate the strengths of various
vertices, $(BK^*\gamma), (\phi\eta\gamma)$, etc. with their
values on the mass-shell.

We have studied the sensitivity of our results on the cut conditions.
To do this, we fix the cut condition to assure the offshellness of the
$\phi$ resonance and then modify the other conditions.
In Table I we present numerical results for representative cuts.
First, 
weak dependence is found for  the hard-photon requirement and  $c$;
with or without these two conditions,
 the decay rates vary within $2\%$
 for both $B_s\to \eta\gamma\gamma$ and $B_s\to \eta'\gamma\gamma$.
Second,
the dependence on the angular cut is moderate;
even using a cut as large as $20^\circ$,
the reductions for the 1PI and the 1PR rates
are within $5\%$,
while  no visible change in the interference effect exists.
Indeed, the interference effect is almost stable against
all different cut conditions.

The insensitivity on the cut conditions
can be understood from two aspects.
First, unlike the quark level transition used in \cite{10}
where the 1PR amplitude is quite singular,
these singularities are absent in the present model.
Second, the new requirement of the offshellness of the
intermediate vector meson eliminates quite a portion of
the phase space for small angle between the 
$P$ and $\gamma$.
Although it is convenient to use also the cut conditions
of \cite{10} in the experimental analyses of the signals,
these conditions are less important on the theoretical side in our case.
 
The only sizable dependence of the decay rates
is on  the cut on the  vector meson resonance.
The  strongest dependence is for the 1PR rates,
which vary almost $20\%$  for $B_s\to \eta\gamma\gamma$
 under the different conditions enforced.
For $B_s\to \eta'\gamma\gamma$ this dependence is much weaker,
since the strong interference of the two possible $\phi$
configurations occurs away from and dominates over the resonant region.
Numerically,
the typical decay rates are
$(1.58\pm 0.18) \times 10^{-20}$GeV for $B_s\to \eta\gamma\gamma$
and $(2.42\pm 0.48) \times 10^{-20}$GeV for $B_s\to \eta'\gamma\gamma$,
if $|\sqrt{m_{\eta^{(\prime)}\gamma}^2}-m_\phi |  \geq 100$MeV
is required.
The errors given here, 
as well as those given for $B\to K\gamma\gamma$,
are due to the uncertainties in the
phases of the interference terms.
Taking $\tau(B_s)=1.61\times 10^{-12}$s,
the corresponding branching ratios are
$(0.39\pm 0.04)\times 10^{-7}$ for $B_s\to \eta\gamma\gamma$
and 
$(0.59\pm 0.12)\times 10^{-7}$ for $B_s\to \eta'\gamma\gamma$,
respectively.
Also, the averaged open angles 
between the two photons for these two channels 
are all around $140^\circ$,
which is stable against the phases of the interference effects
and is comparable to the value $\sim 135^\circ$ in \cite{10}.
%The distribution of this angle 
%
For $B\to K\gamma\gamma$,
again,
the only sizable dependence of the decay rates
is on  the cut on the  vector meson resonance.
The decay rates are
$(2.87\pm 0.39)\times 10^{-20}$Gev for $B^0\to K^0\gamma\gamma$ and
$(2.04\pm 0.26)\times 10^{-20}$Gev for $B^\pm\to K^\pm\gamma\gamma$,
if $|\sqrt{m_{K\gamma}^2}-m_{K^*} |  \geq 300$MeV is taken.
They correspond to branching ratios of
$(0.68\pm 0.09)\times 10^{-7}$ for $B^0\to K^0\gamma\gamma$ and
$(0.50\pm 0.06)\times 10^{-7}$ for $B^\pm\to K^\pm\gamma\gamma$.
The averaged open angle between the two $\gamma$'s is
also around $140^\circ$.

To summarize,
our results reveal interesting features of the exclusive decays
 $B_s\to\eta(\eta')\gamma\gamma$.
Firstly,
inspection of Table I indicates that the short-distance(1PI)
and the long-distance(1PR) components play
a comparable role in these decays,
which is a new feature in the domain
of $B$ radiative decays.

Secondly,
the calculated exclusive modes appear to constitute a 
very sizable portion of the respective inclusive decays.
For example,
in \cite{10} it was found 
$BR(B\to X_s\gamma\gamma)\sim 1\times 10^{-7}$,
to which our results for $B\to K\gamma\gamma$ are compared.
We remark, however, 
that unlike the $B\to X_s\gamma$ transition where
the dominant mechanism is the quark level $b\to s\gamma$ transition,
in the case of $B\to X_s\gamma\gamma$ the diagram with one photon
attached to the spectator quark also contributes.
This mechanism is accounted by our treatment but
has not been included in \cite{10},
which might affect the ratios of the calculated exclusive
and inclusive decays.

 We would like to acknowledge discussions with Prof. G. Eilam.
This research  is supported in part by Grant 5421-3-96
from the Ministry of Science and the Arts of Israel. 
The research of   P.S. has also been supported in part 
by the Fund for Promotion of Research at the Technion.
%\newpage

\newpage
\small

\noindent {\bf TABLE I.} 
Decay rate for $B_s\to \eta\gamma\gamma$,
$B_s\to \eta'\gamma\gamma$,
$B^0\to K^0\gamma\gamma$
and $B^\pm\to K^\pm\gamma\gamma$
 (in $10^{-20}$GeV).\\[3mm]
\begin{tabular}{c|ccc|ccc}
\hline\hline
& &no angular cut& & &all angles $\geq 20^\circ$& \\
Res. Cut&  1PI &1PR & Inter. &  1PI &1PR & Inter.\\
\hline\hline
$B_s\to \eta\gamma\gamma$ &&&&&&\\
\hline
$|\sqrt{m_{\eta\gamma}^2}-m_\phi | \geq 50$MeV  
& 0.95 & 0.83 & 0.18 & 0.92 & 0.79 & 0.18
\\
$|\sqrt{m_{\eta\gamma}^2}-m_\phi |  \geq 100$MeV
& 0.92 & 0.66 & 0.18 & 0.90 & 0.64   & 0.18
\\
%  \hline
%  $\sqrt{m_{\eta\gamma}^2}\geq 1.1$GeV 
%  & 0.93 & 0.73 & 0.18 & 0.91 & 0.70  & 0.18
%  \\
$\sqrt{m_{\eta\gamma}^2}\geq 1.2$GeV 	          
& 0.87 & 0.56 & 0.17 & 0.87 & 0.55 & 0.17
\\
\hline\hline
$B_s\to \eta'\gamma\gamma$ &&&&&&\\
\hline
$|\sqrt{m_{\eta'\gamma}^2}-m_\phi | \geq 50$MeV  
& 1.77 & 0.66 & 0.48 & 1.72 & 0.63 & 0.46 
\\
$|\sqrt{m_{\eta'\gamma}^2}-m_\phi |  \geq 100$MeV
& 1.77 & 0.65 & 0.48 & 1.72 & 0.62 & 0.46
\\
%  \hline
%  $\sqrt{m_{\eta'\gamma}^2}\geq 1.1$GeV 
%  & 1.77 & 0.65 & 0.48 & 1.72 & 0.62 & 0.46
%  \\
$\sqrt{m_{\eta'\gamma}^2}\geq 1.2$GeV 	          
& 1.77 & 0.64 & 0.47 & 1.72 & 0.61 & 0.46
\\
\hline\hline
 $B^0\to K^0\gamma\gamma$ &&&&&&\\
\hline
$|\sqrt{m_{K\gamma}^2}-m_{K^*} | \geq 300$MeV  
& 1.40& 1.48   &  0.39  
& 1.40  & 1.47   &  0.39  
\\
$|\sqrt{m_{K\gamma}^2}-m_{K^*} |  \geq 500$MeV
& 1.31  & 1.30   &  0.37  
& 1.31  & 1.29   &  0.37  
\\
%  \hline
%  $\sqrt{m_{K\gamma}^2}\geq 1.2$GeV 
%  & 1.38  & 1.43  &  0.38  
%  & 1.38  & 1.41  &  0.38  
%  \\
$\sqrt{m_{K\gamma}^2}\geq 1.4$GeV 	          
& 1.24   & 1.19   &  0.37  
&  1.24  & 1.19   &  0.37  
\\
\hline\hline
 $B^\pm\to K^\pm\gamma\gamma$ &&&&&&\\
\hline
$|\sqrt{m_{K\gamma}^2}-m_{K^*} | \geq 300$MeV  
&  1.40 &   0.64 &   0.26 
&  1.40 &   0.64 &   0.26 
\\
$|\sqrt{m_{K\gamma}^2}-m_{K^*} |  \geq 500$MeV
&  1.31 &   0.56 &   0.25 
&  1.31 &   0.56 &   0.25 
\\
%  \hline
%  $\sqrt{m_{K\gamma}^2}\geq 1.2$GeV 
%  &  1.38 &  0.62 &   0.25 
%  &  1.38 &  0.61 &   0.25 
%  \\
$\sqrt{m_{K\gamma}^2}\geq 1.4$GeV 	          
&   1.24 &   0.52 &   0.24 
&   1.24 &   0.51 &   0.24 
\\
\hline\hline
\end{tabular}
\end{document}